\documentclass[prd,aps,preprint,tightenlines,showpacs,nofootinbib,superscriptaddress]{revtex4-1}
     \usepackage{mathrsfs}
     \usepackage{amsfonts}
     \usepackage{amsmath}
     \usepackage{array}
     \usepackage{verbatim}
     \usepackage{bm}
     \usepackage{epsfig}
     \usepackage{graphicx,color}
     \usepackage{relsize}
    \usepackage{lineno}
     \usepackage{multirow}
     \usepackage[utf8]{inputenc}
     \RequirePackage{xspace}

\begin{document}
	     	\author{Ya-Ping Xie}\email{xieyaping@impcas.ac.cn}
     	\affiliation{Institute of Modern Physics, Chinese Academy of
     		Sciences, Lanzhou 730000, China}
\affiliation{ University of Chinese Academy of Sciences, Beijing 100049, China}
      	\author{Xu Cao}\email{caoxu@impcas.ac.cn}
      	\affiliation{Institute of Modern Physics, Chinese Academy of
      		Sciences, Lanzhou 730000, China}
\affiliation{ University of Chinese Academy of Sciences, Beijing 100049, China}
      	\author{Yu-Tie Liang}\email{ liangyt@impcas.ac.cn}
      	\affiliation{Institute of Modern Physics, Chinese Academy of
      		Sciences, Lanzhou 730000, China}
\affiliation{ University of Chinese Academy of Sciences, Beijing 100049, China}
     	\author{Xurong Chen}\email{xchen@impcas.ac.cn}
     	\affiliation{Institute of Modern Physics, Chinese Academy of
     		Sciences, Lanzhou 730000, China}
\affiliation{ University of Chinese Academy of Sciences, Beijing 100049, China}
\affiliation{ Institute of Quantum Matter, South China Normal University, Guangzhou 510006, China}

     	\title{Production of hidden-charm and hidden-bottom pentaquark states in electron-proton collisions}
\begin{abstract}
Electro-production of several pentaquark states is investigated in this paper.
eSTARlight package is adapted to study the electro-production of $J/\psi$ and $\Upsilon (1S)$ via pentaquark $P_c$ and $P_b$ resonance channels in $e p \to e J/\psi p$ and $e p \to e\Upsilon(1S) p$ scattering processes at proposed electron-ion colliders (EICs). The results in this paper are compared to the non-resonance $t$-channels, which is described in pomeron exchange model in our studies. Some pseudo-rapidity distributions rapidity distributions of $J/\psi$  and $\Upsilon(1S)$ are presented for proposed EICs including EicC and EIC-US. It is found that EicC is a good platform to identify $P_b$ states in the future.

  \end{abstract}
  \maketitle
  \section{introduction}
Up to now a rich spectrum of the exotic mesons, including charmonium-like and bottomonium-like states, are emerging, and more new states are expected for the continuing experimental effort~\cite{Chen:2016qju,Diehl:2003qa,Guo:2017jvc,Lebed:2016hpi,Esposito:2016noz,Olsen:2017bmm,Liu:2019zoy,Wang:2019got,Brambilla:2019esw,Ali:2017jda}.
However, in the baryon sector only three narrow pentaquark states, $P_c(4312)$, $P_c(4440)$ and $P_c(4457)$, are discovered by the LHCb collaboration in $\Lambda_{b}\to J/\psi pK^-$ decay~\cite{Aaij:2015tga,Aaij:2019vzc}.
It is essential to study these known states and search for new states by other decay and reaction channels in order to disentangle different models. Just recently, D0 and GlueX collaborations have searched for these states in inclusive $p \overline p $ collisions~\cite{Abazov:2019kwn} and photoproduction~\cite{Ali:2019lzf}, recpectively. The D0 collaboration found an enhancement from joint contribution of $P_c(4440)$ and $P_c(4457)$ in $J/\psi p$ invariance mass spectrum with low significance~\cite{Abazov:2019kwn}, serving as the first and only confirmatory evidence for these pentaquark states.

Various interpretations were proposed for the nature of hidden charm pentaquark states before and after their observation, e.g. molecular states~\cite{Wu:2010jy,Wu:2010vk}, compact diquark-diquark-antiquark states~\cite{Cheng:2019obk,Ali:2019npk,Ali:2019clg}, and hadro-charmonium states~\cite{Eides:2019tgv}.
In addition, it is pointed out that the peaks of pentaquark in the decay and reaction with multi-particle final states could be induced by triangle singularity considering that their masses locate close to the $\Sigma_c\bar{D}$ and $\Sigma_c\bar{D}^*$ threshold~\cite{Guo:2015umn,Liu:2015fea,Liu:2016dli,Guo:2016bkl,Bayar:2016ftu,Liu:2019dqc,Guo:2019twa}.
In order to survey this non-resonance explanation, the reactions with two-body final states induced by beams of photon, electron~\cite{Wang:2015jsa,Karliner:2015voa,Kubarovsky:2015aaa,Huang:2016tcr,Blin:2016dlf,Wu:2019adv,Wang:2019krd} and pion~\cite{Lu:2015fva,Liu:2016dli,Wang:2019dsi,Kim:2016cxr} are suggested to be decisive. At present and in the near future, the high energy pion beam seems to be unavailable, so photo- and electroproduction reactions would play the central role and attract much interest. These reactions are also useful to search for other $P_c$, for instance those among seven states in spin multiplets anticipated by heavy-quark spin symmetry~\cite{Liu:2019tjn,Xiao:2019aya,Du:2019pij}, and also $P_b$, the bottom analogs of $P_c$, expected by heavy quark flavor symmetry in many models~\cite{Wu:2010rv,Xiao:2013jla,Karliner:2015voa,Karliner:2015ina}.

The observation of hidden-charm pentaquark states encourage people to investigate the hidden-bottom pentaquark state which contains a bottom quark pair and three light quarks. There are several  papers about investigation for the nature of hidden-bottom pentaquark state in different models\cite{Wu:2017weo,Huang:2018wed,Gutsche:2019mkg}. Photoproduction of hidden-bottom pentaquark state has been investigated in Refs.\cite{Wang:2019zaw,Cao:2019gqo}. It is natural to predict the production of hidden-bottom pentaquark state in electron-proton scattering
in future EICs.

Electron-Ions Collider (EIC) is an important platform to explore nuclear structure and exotic particle nature in next decade. In electron-proton scattering, the initial electron emits virtual photon which interacts with the initial proton to produce vector mesons. There are several proposed EICs, for instance, EicC (EIC in China)~\cite{CAO:2020EicC,Chen:2018wyz}, EIC-US (EIC in US) \cite{Morozov:2017} and LHeC (EIC in LHC) \cite{AbelleiraFernandez:2012cc}, ranging from intermediate to extremely high energies.  

The simulation work of production in EICs is very important before EICs are built. The simulation work can help us to estimate the particles cross sections for the proposed EICs.
eSTARlight is a Monte-Carlo package to simulate production of vector mesons in electron-proton scattering for EICs \cite{Lomnitz:2018juf}. It can de describe the vector meson production well of HERA in the $t$-channel.  The production of exotic particles were also studied in eSTARlight\cite{Klein:2019avl}. The cross sections of photon-proton to vector mesons
is necessary to calculate the cross section of vector mesons in electron-proton scattering
In eSTARlight package, the glauber model is employed to obtain the $\gamma A\to VA$\cite{Klein:1999qj}.
 With the helps of eSTARlight, we can obtain the four momentum of final state particles.which are important for the detector systems. Then we can rebuild the four momentum of short-life particles. The simulation can provide some distributions of 
the physical process. In previous versions of eSTARlight, only the $t$-channel
 is investigated. In this work, we are going to study the $s$-channel vector meson production in electron-proton scattering using  eSTARlight.
We investigate the electroproduction of pentaquark $P_c$ in $e p\to e J/\psi p$ and $P_b$ in $e p\to e \Upsilon(1S) p$ scattering with a great detail in this paper. Here we will concentrate on EicC and EIC-US by comparison of cross sections and the rapidity distributions of final particles.

The main aim of this paper is adopting eSTARlight to simulate the production of charm and bottom vector mesons in the $s$-channel and $t$-channel.  eSTARlight can describe vector mesons cross sections of HERA well in the $t$-channel. We extend the vector mesons production in the $s$-channel in eSTARlight.
This paper is organized as follows. The theoretical framework is given in Sec~\ref{sec:framework}. The numerical results are presented in Sec.~\ref{sec:numerical}, closed with a summary in Sec.~\ref{sec:conclusion}.

 \section{Theoretical Framework} \label{sec:framework}
\begin{figure}
	\centering
\includegraphics[width=0.8\textwidth]{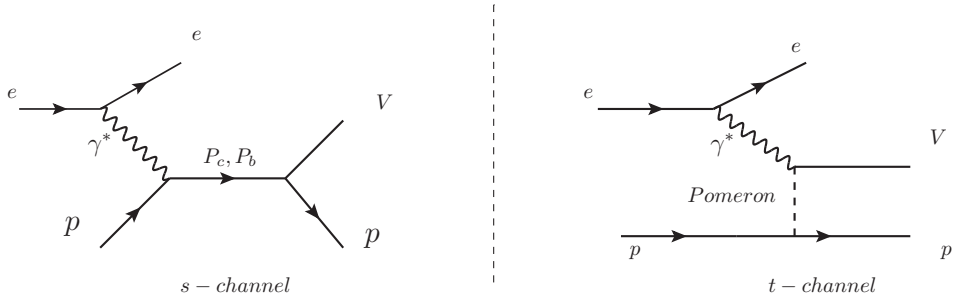}
	\caption{ Diagrams for $J/\psi$ and $\Upsilon (1S)$ production in electron-proton scattering via $P_c$ and $P_b$ pentaquark resonances exchange $s$-channel (left graph) and pomeron exchange $t$-channel (right graph). }
	\label{eppc}
\end{figure}

In electron-proton scattering, diffractive production of vector meson is important since the photon in electro-production is
off-mass-shell. It is interesting to see how the internal structures of the particles involved influence the vector mesons 
production in electron-proton scattering.
The diagrams for $s$-channel and $t$-channel of $ep\to eVp$ are depicted in Fig.~\ref{eppc}. In the $s$-channel (left graph), the virtual photon and initial proton produce resonances (e.g. $P_c$ and $P_b$ states), and then the pentaquark resonance states decay into vector mesons and proton. In the $t$-channel(right graph), the virtual photon interacts with proton via exchanging pomerons or gluons and then converts into final vector mesons. In this paper we use the pomerons exchanging in the $t$-channel. We treat the $t$-channel contribution as a background of pentaquark states resonance contributions. We parameterize the cross section of $\gamma p\to Vp$, as the basic input to the simulation of $ep\to eVp$ reaction. This can be recognized by the eSTARlight package. 

In the electron proton scattering, the cross section of  $ep\to eVp$ are in terms of the cross section of $\gamma^* p\to Vp$. It is written as  \cite{Lomnitz:2018juf},
     	\begin{eqnarray}
     	\sigma(ep\to eVp)=\int dkdQ^2\frac{dN^2(k,Q^2)}{dkdQ^2}\sigma_{\gamma^* p\to Vp}(W,Q^2).
   \end{eqnarray}
where $k$ is the momentum of the photon emitted from initial electron in the target rest frame, W is the center of mass (c.m.) energy of the virtual photon and proton system, and $Q^2$ is the virtuality of the virtual photon. The photon flux is given as \cite{Budnev:1974de}
\begin{eqnarray} \label{eq:photoflux}
\frac{d^2N(k,Q^2)}{dkdQ^2}=\frac{\alpha}{\pi kQ^2}\Big[1-\frac{k}{E_e}+\frac{k^2}{2E^2_e}-\Big(1-\frac{k}{E_e}\Big)\Big|\frac{Q^2_{min}}{Q^2}\Big|\Big].
\end{eqnarray}
where $E_e$ is the energy of the incoming electron in the proton rest frame, and $Q^2_{min}$ is defined as
\begin{eqnarray}
Q^2_{min}=\frac{m_e^2k^2}{E_e(E_e-k)}.
\end{eqnarray}
The maximum $Q^2$ is determined by the energy loss of the initial electron, it reads
\begin{eqnarray}
Q^2_{max}=4E_e(E_e-k).
\end{eqnarray}
The $Q^2$ dependence of $\sigma_{\gamma^*p\to Vp}(W,Q^2) $ is factorized as
\begin{eqnarray} \label{eq:Q2xsection}
\sigma_{\gamma^*p\to Vp}(W,Q^2)&=&\sigma_{\gamma p\to Vp}(W,Q^2=0)\bigg(\frac{M_V^2}{M_V^2+Q^2}\bigg)^\eta.
\end{eqnarray}
where $\eta=c_1+c_2(M_V^2+Q^2)$ with the values of $c_1 = 2.36\pm0.20$ and $c_2 =0.0029\pm 0.43\!\!\!\!\!\quad\mathrm{ GeV}^2$, which are determined by the data of $\gamma^*p\to Vp$ with $Q^2 \neq 0$ \cite{Lomnitz:2018juf}. We use the same $Q^2$ dependence for pentaquark and pomeron channels, as these values are unknown for pentaquark resonance channel. Because of the very strong $Q^2$ dependence of photon flux in Eq.~(\ref{eq:photoflux}), the impact of this prescription is expected to be not big for the final results.

For the pentaquark states resonance channel, the cross sections of $\gamma p{\to}Vp$ can be written in a compact Breit-Wigner form\cite{Kubarovsky:2015aaa,Karliner:2015voa}
\begin{eqnarray}
\sigma^{P_X}_{\gamma p{\to}Vp}(W)=\frac{2J+1}{2(2s_2+1)}\frac{4\pi}{k^2_{in}}\frac{\Gamma^2_{P_X}}{4}
\frac{\mathcal{B}(P_X\to\gamma p)\mathcal{B}(P_X\to Vp)}{(W-M_{P_X})^2+\Gamma^2_{P_X}/4}.
\end{eqnarray}
where $P_X$ denotes pentaquark states, such $P_c$ and $P_b$.  $s_1$ is the spin of initial proton and $J$ is the total spin of $P_c$ and $P_b$ pentaquark states. Here $M_{P_X}$ and $\Gamma_{P_X}$ is the mass and total decay width of the $P_c$ and $P_b$ states, respectively. The $k_{in}$ is the magnitude of three momentum of initial state in the c.m. frame.  The branching ratio of $P_X\to \gamma p$ is calculated by the vector meson dominant model:
\begin{eqnarray}
\mathcal{B}(P_X\to \gamma p)=\frac{3\Gamma(V \to e^+e^-)}{\alpha M_V}
\Big(\frac{k_{in}}{k_{out}}\Big)^{2L+1}\mathcal{B}(P_X\to Vp).
\end{eqnarray}
where $\alpha$ is the fine structure constants and $\Gamma (V\to e^+e^-)$ is the dilepton decay width of vector mesons.
The $k_{out}$ is the magnitude of three momentum of final state in the c.m. frame. In this work, we use the lowest orbital excitation $L=0$ for $J/\psi+ p$ system and $J=1/2$. Other quantum numbers of $P_X$ can be similarly calculated. We adopt $\mathcal{B}(P_c\to J/ \psi p) = 5\%$ and $\mathcal{B}(P_b\to \Upsilon(1S)p) = 5\%$ for the calculations in this work, which are in the same level of the upper limits from GlueX group~\cite{Ali:2019lzf}. A comparison of our $\sigma^{P_c}_{\gamma p{\to}J/\psi p}(W)$ to the GlueX data could be found in Ref.~\cite{Cao:2019kst}.

In order to study the rapidity distributions and transverse momentum distributions of vector mesons and proton in final states, we need angular distributions of the decay process $P_X \to Vp$.
In the process of $P_X\to Vp$, the angle distribution of $P_X\to Vp$ has following general expression
\begin{eqnarray}
\frac{d\sigma}{d\cos\theta}\propto 1+\beta \cos^2\theta.
\end{eqnarray}
Here $\theta$ is polar angle of vector meson or proton in the rest frame of $P_c$ and $P_b$ states and $\beta$ is dependent on the quantum number $J^p$ of $P_X$ pentaquark, if only lowest partial wave is considered. But usually several partial waves are presented in this work, so the actual value of $\beta$ would deviate from these values.  The relation of $\beta$ and $J^p$ are listed in Table.\ref{table00}. These results are employed in the calculation of $J/\psi$ and $\Upsilon(1S)$ rapidity distributions
\begin{table}[!h]
\begin{tabular}{|p{1cm}<{\centering}|p{1cm}<{\centering}|p{1cm}<{\centering}|p{1cm}<{\centering}|p{1cm}<{\centering}|}
\hline
\hline
$J^p$  &  $\frac{1}{2}^-$ & $\frac{1}{2}^+$    &    $\frac{3}{2}^-$ & $\frac{3}{2}^+$ \\
\hline
$\beta$ &-1 &0 & 0 & 1\\
\hline
\hline
\end{tabular}\caption{$\beta $ from different quantum number of $P_c$ and $P_b$ states.  }
	\label{table00}
\end{table}

For the contribution of Pomeron exchange $t$-channel, the cross section of $\gamma p\to Vp$ is given as \cite{Klein:2016yzr},
\begin{eqnarray}
\sigma^t_{\gamma p\to Vp}(W)=\sigma_p\cdot\Big(1-\frac{(m_p+m_{V})^2}{W^2}\Big)\cdot W^\epsilon,
\end{eqnarray}
with $\sigma_p$ = 4.06 nb and $\epsilon$ = 0.65 for $J/\psi$ and $\sigma_p$ = 6.4 pb and $\epsilon$ = 0.74 for $\Upsilon(1S)$, which are determined by the experimental data of $\gamma p\to Vp$ with $Q^2 = 0$ and applied successfully to previous studies of $J/\psi$ and $\Upsilon(1S)$ electroproduction \cite{Klein:2016yzr}.

In this work, we employ eSTARlight to simulate pentaquark states resonance production processes via photon-proton interaction at first. Then, the decay process of $P_c\to J/\psi+p$ and $P_b\to \Upsilon(1S)+p$ are implemented in eSTARlight. Finally, the vector mesons to dilepton is simulated.  The resonance channel production in eSTARlight is newly studied and it can be applied to considered other resonance channel in the next step.

 \section{Numerical result} \label{sec:numerical}
In this work, two pentaquark states $P_c(4312)$ and $P_b(11120)$ are selected to study the vector mesons production.
The properties of $P_c(4312)$ and $P_b(11120)$ are listed in Table.~\ref{table01}, where the decay width of $P_b(11120)$ is
taken from Ref.\cite{Cao:2019gqo}. Throughout this paper we use the central values of the masses of two pentaquark states. We investigate their production in proposed EICs, including EicC and EIC-US, whose collider energies are listed. A detailed comparison of the proposed EICs are presented in Ref. \cite{CAO:2020EicC,Klein:2019avl}.

First of all, we present the estimated $J/\psi$ and $\Upsilon(1S)$ cross sections of in the $s$-channels and $t$-channel in Table.~\ref{table01}. 
The cross sections of the $t$-channel is viewed as the background of the $t$-channel pentaquark production. 
For all the calculation in this paper, $0<Q^2<5$ $\mathrm{GeV}^2$  and $\beta =-1$ are employed. From Table.~\ref{table01}, it implies that
the $J/\psi$ cross section in the $t$-channel is much larger than the $s$-channel in both EicC and EIC-US. However, the cross sections of $\Upsilon(1S)$
in the $t$-channel are not so much larger than the $s$-channel as $J/\psi$ production. This conclusion is very important for the studying the pentaquark states 
because the $t$-channel can be viewed as a background for identifying pentaquark states in experiments. 

\begin{figure}[h]
	\centering
\includegraphics[width=0.45\textwidth]{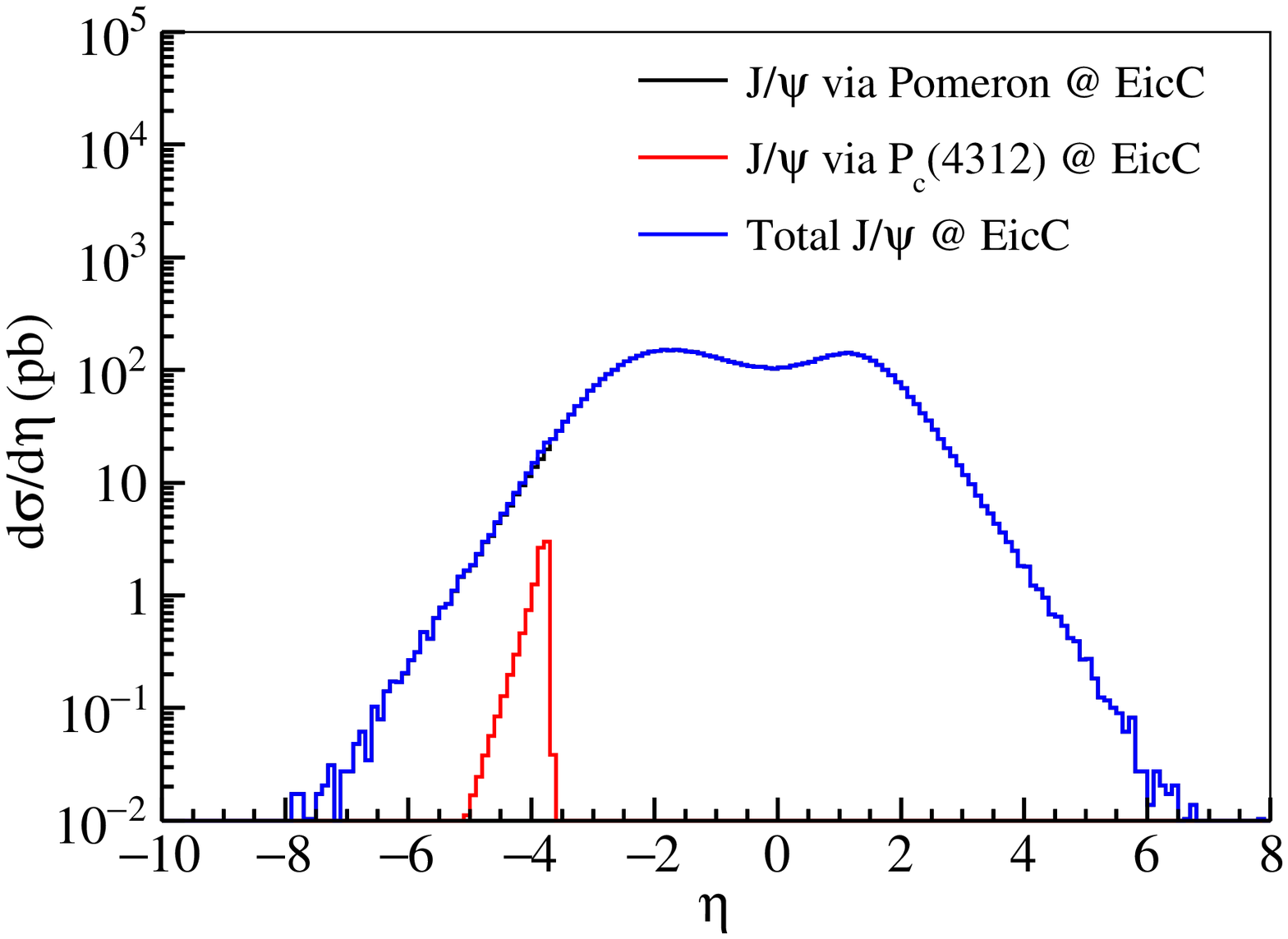}
\includegraphics[width=0.45\textwidth]{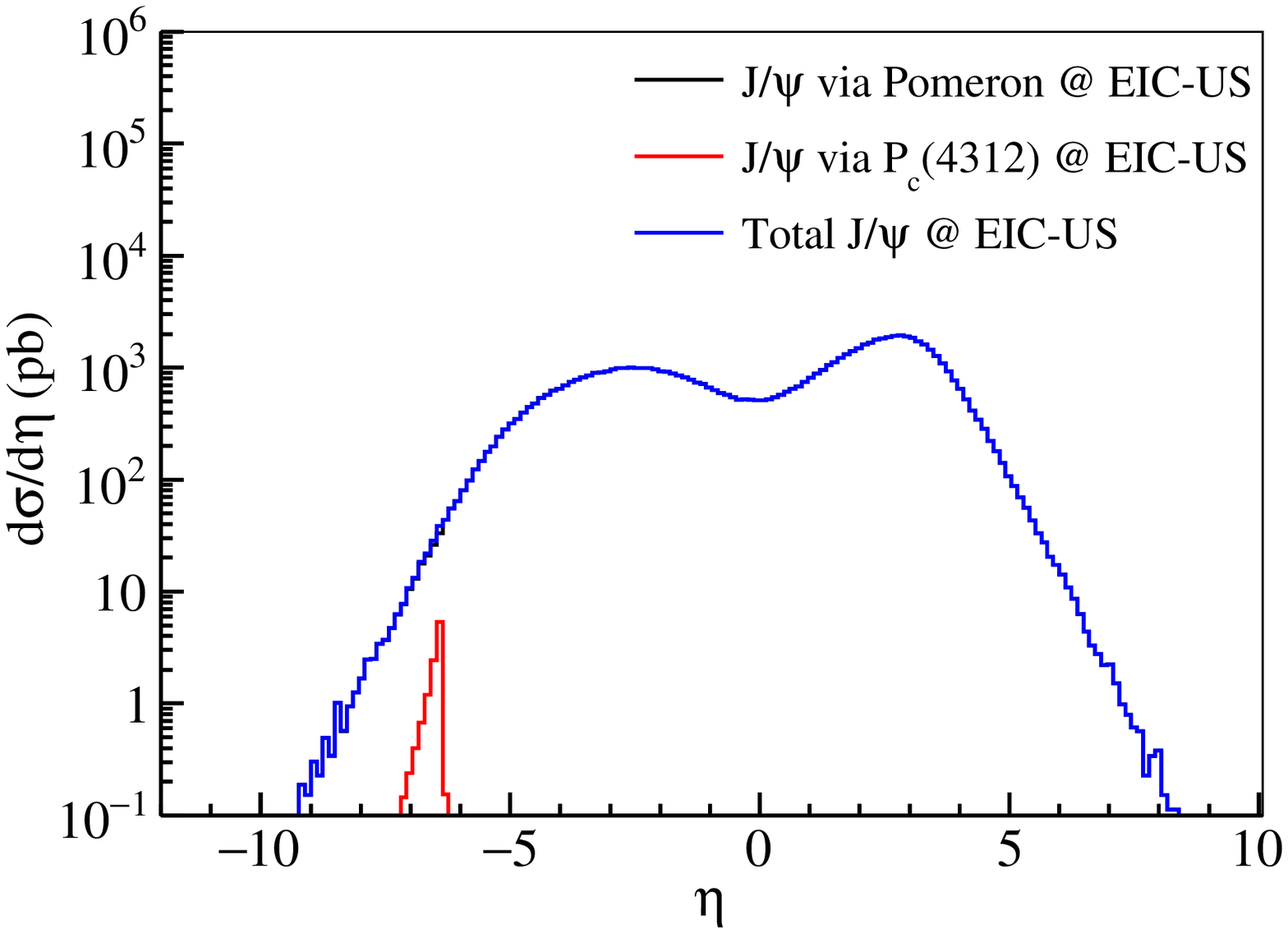}
	\caption{(Color online) Pseudo-rapidity distributions of $J/psi$ in two channels for EicC (left graph) and EIC-US (right graph). }
	\label{fig01}
\end{figure}

\begin{table}[h]
\begin{tabular}{|c|c|c||c|c|c|}
\hline
\hline
\multirow{2}{*} {States}&\multirow{2}{*}&\multirow{2}{*}{Properties~\cite{Aaij:2019vzc,Cao:2019gqo}}  &Collider & EicC & EIC-US\\
\cline{4-6} & &&  Energy ( e.vs. p)&3.5 GeV vs 20 GeV & 18 GeV vs 275 GeV  \\
\cline{1-3}\cline{4-6}
\multirow{2}{*}{ $P_c(4312)$}&Mass&$4.311\pm0.7^{+6.8}_{-0.6}$ GeV &$\sigma_t(ep\to eJ/\psi p)$ &0.69 nb& 9.1 nb\\
\cline{2-6}
&Width& $9.8\pm2.7^{+3.7}_{-4.5}$ MeV &$\sigma_s(ep\to eJ/\psi p)$ &0.89 pb  &1.3 pb \\
\hline

\multirow{2}{*}{$P_b(11120)$}&Mass&11.120 GeV &$\sigma_t(ep\to e\Upsilon p)$&0.13 pb& 15 pb\\
\cline{2-6}
&Width&30 -- 300 MeV &$\sigma_s(ep\to e\Upsilon p)$ &9.3 -- 82 fb& 0.022 --0.19 pb \\
\hline
\hline
\end{tabular}\caption{Cross sections $J/\psi$ and $\Upsilon(1S)$ vector mesons in two channel for proposed EicC and EIC-US. The $s$-channel
is the pentaquark states resonance channels.  }
	\label{table01}
\end{table}
Secondly, We present the pseudo-rapidity distributions of $J/\psi$ in two channels for proposed EicC and EIC-US in Fig.~\ref{fig01}.
Since the cross section of the $t$-channel of $J/\psi$ is much larger than the $s$-channel, it can be seen that the 
$s$-channel cross section is smaller than the $t$-channel. Consequently, we can neglect the interference between the $t$-channel and the $s$-channel because the amplitude in the $s$-channel is much small than the $t$-channel. 

\begin{figure}[h]
	\centering
\includegraphics[width=0.45\textwidth]{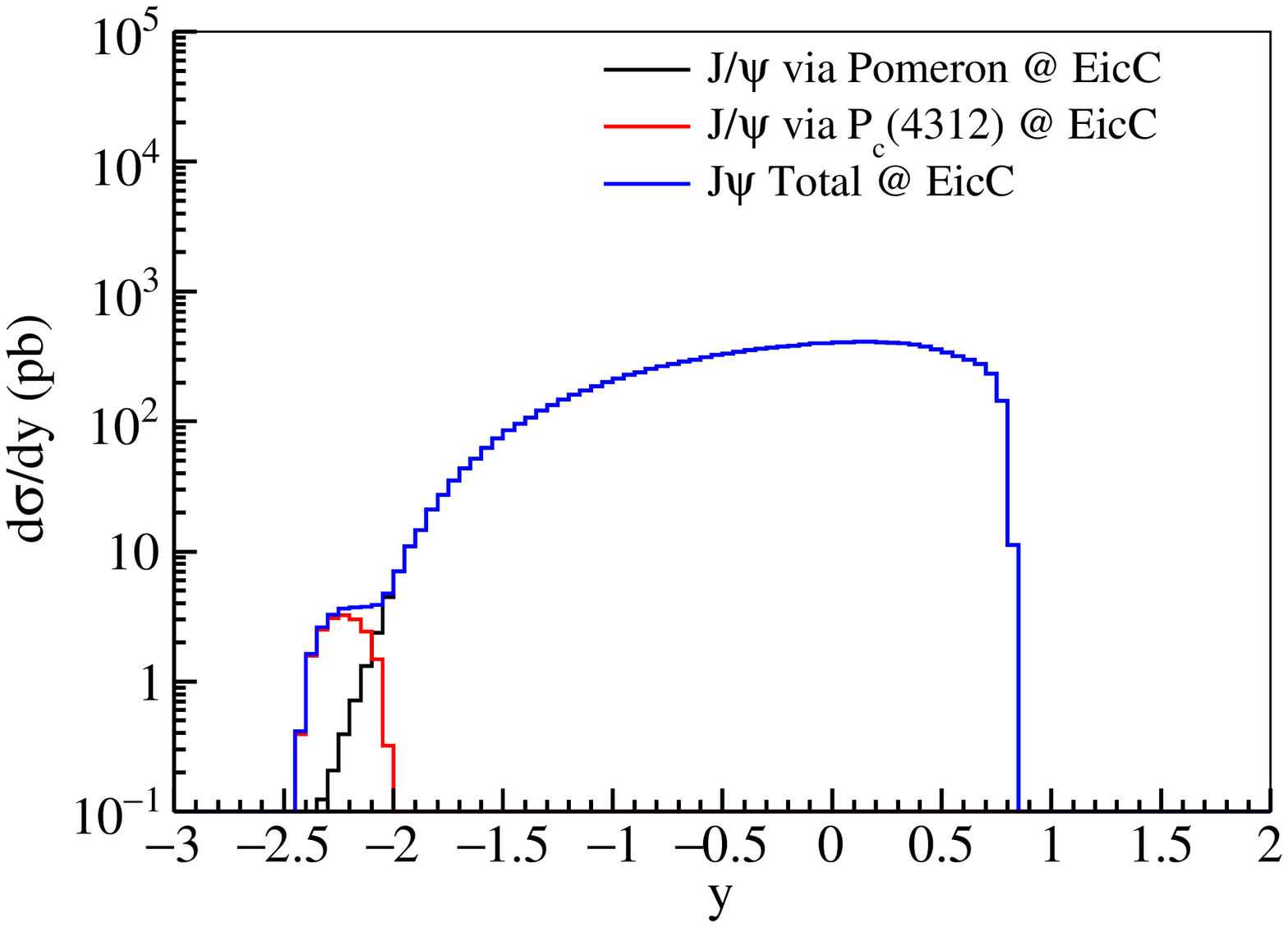}
\includegraphics[width=0.45\textwidth]{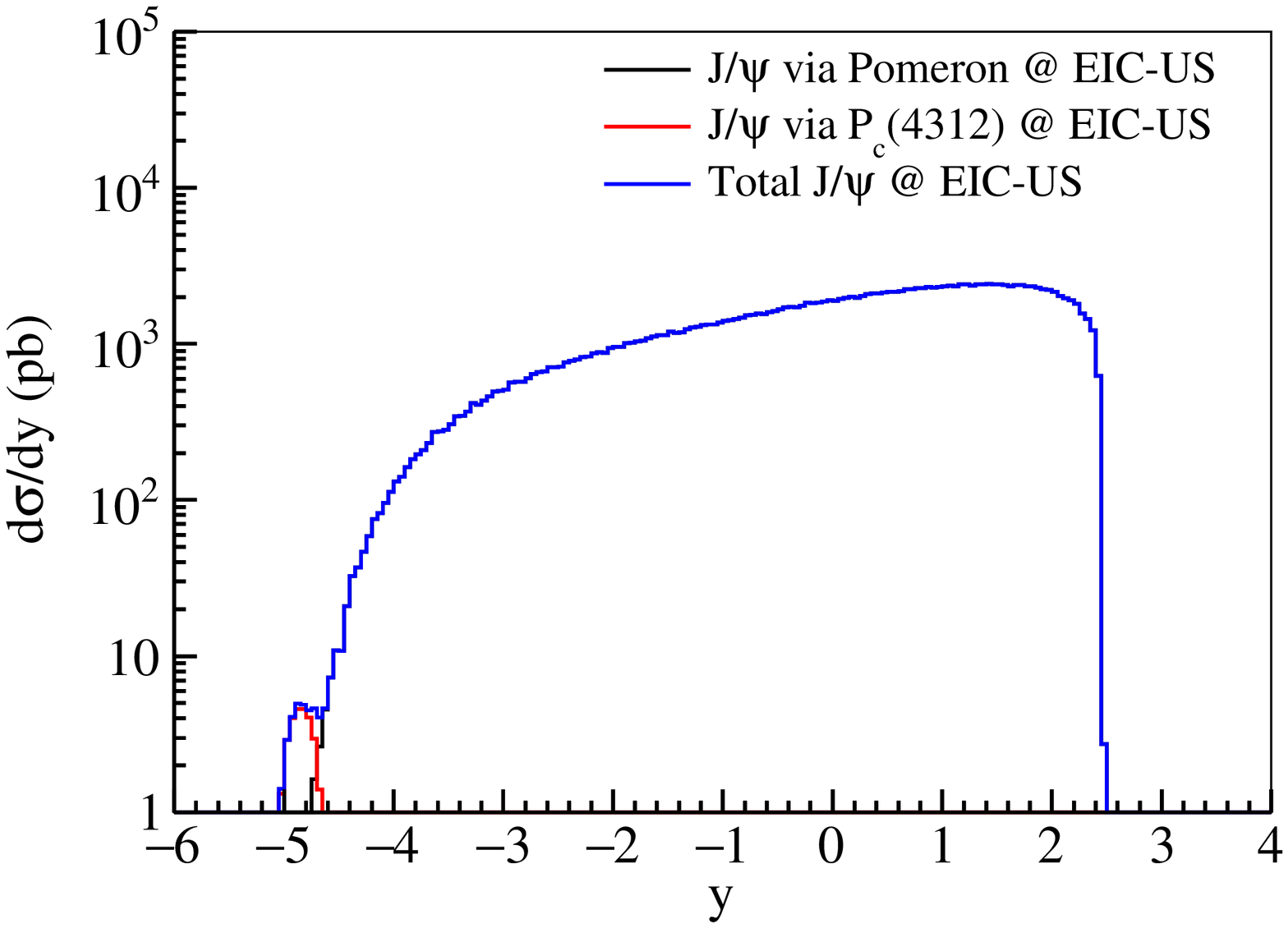}
	\caption{(Color online) Rapidity distributions of $J/\psi$ produced in two channels for proposed EicC (left graph) and EIC-US (right graph). }
	\label{fig02}
\end{figure}

\begin{figure}[h]
	\centering
\includegraphics[width=0.45\textwidth]{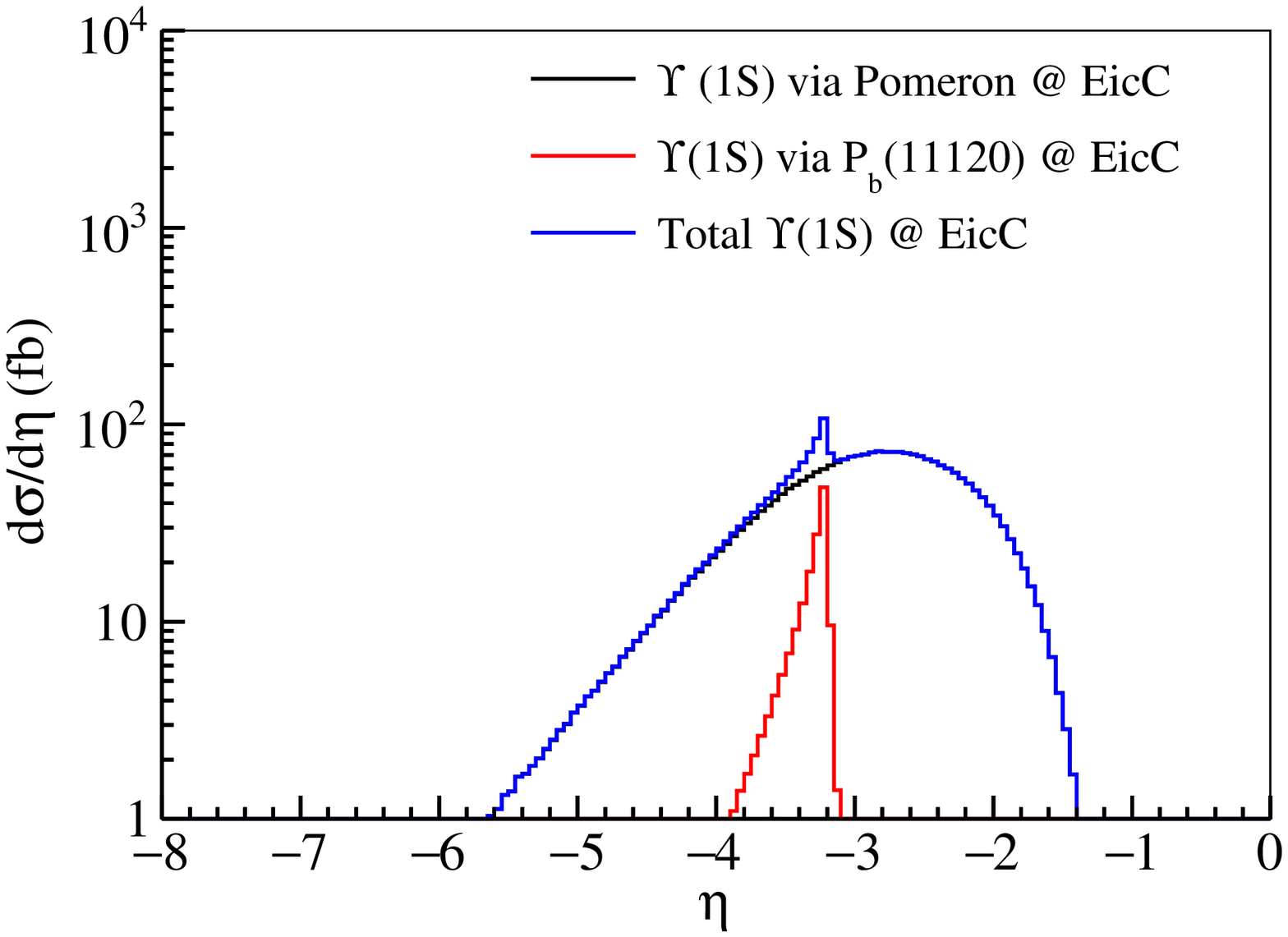}
\includegraphics[width=0.45\textwidth]{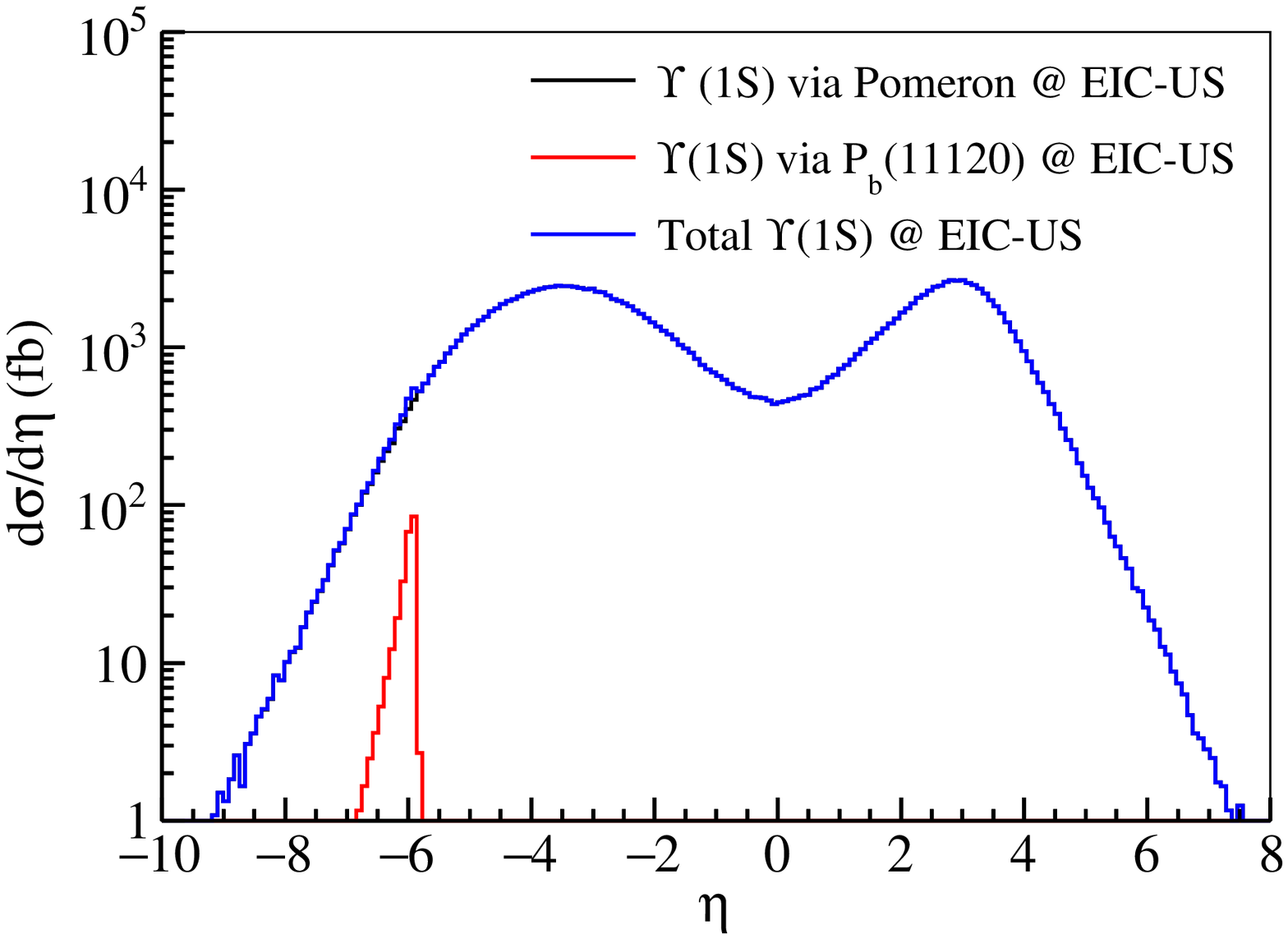}
	\caption{(Color online) pseudo-rapidity distributions of $\Upsilon(1S)$ in two channels for proposed EicC (left graph) and EIC-US (right graph). The
	width of $P_b(11120)$ 30 MeV is taken in the calculations.}
	\label{fig03}
\end{figure}

\begin{figure}[h]
	\centering
	\includegraphics[width=0.45\textwidth]{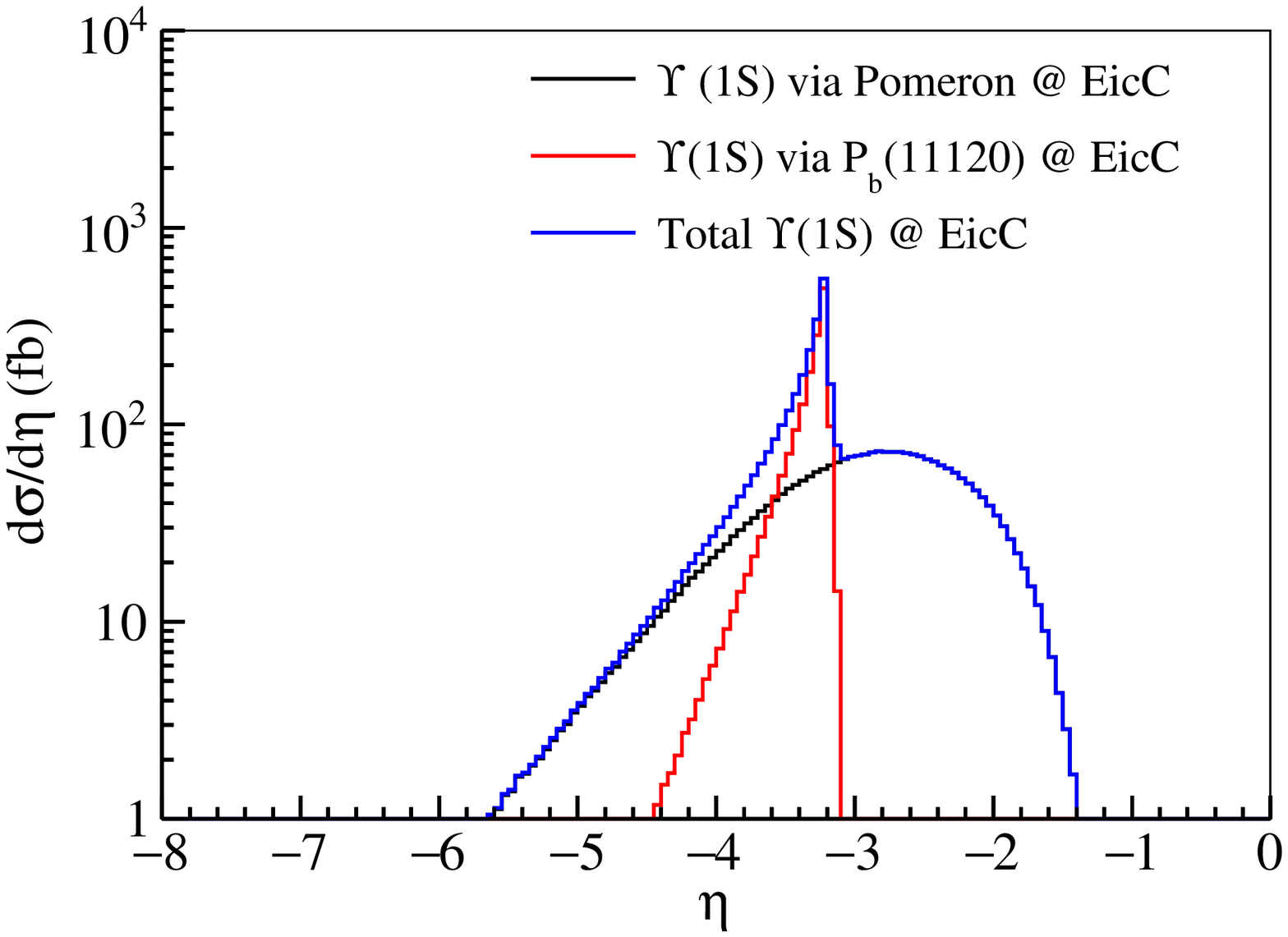}
	\includegraphics[width=0.45\textwidth]{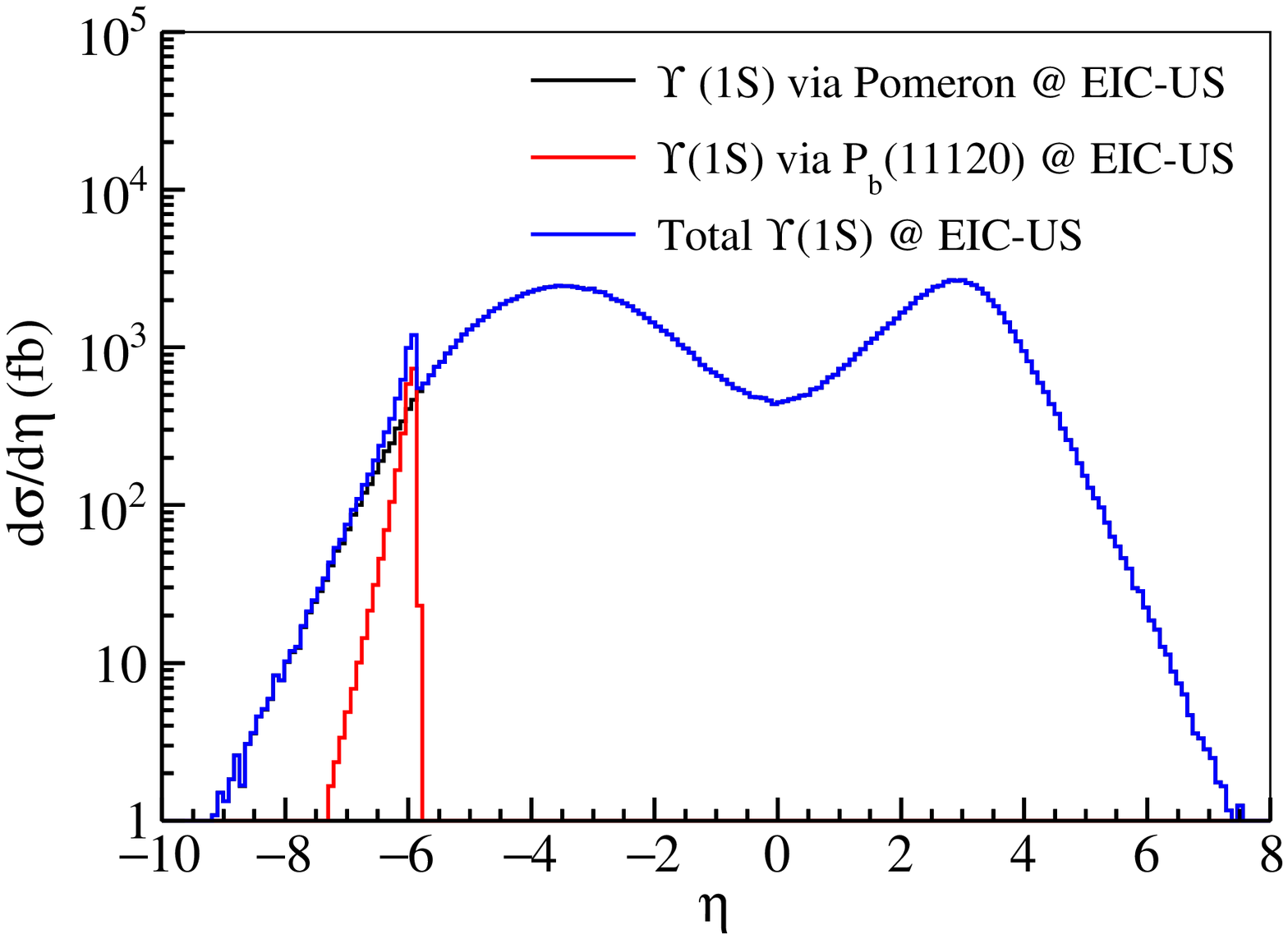}
	\caption{(Color online) pseudo-rapidity distributions of $\Upsilon(1S)$ in two channels for proposed EicC (left graph) and EIC-US (right graph). The
		width of $P_b(11120)$ 300 MeV is taken in the calculations.}
	\label{fig04}
\end{figure}

\begin{figure}[h]
	\centering
\includegraphics[width=0.45\textwidth]{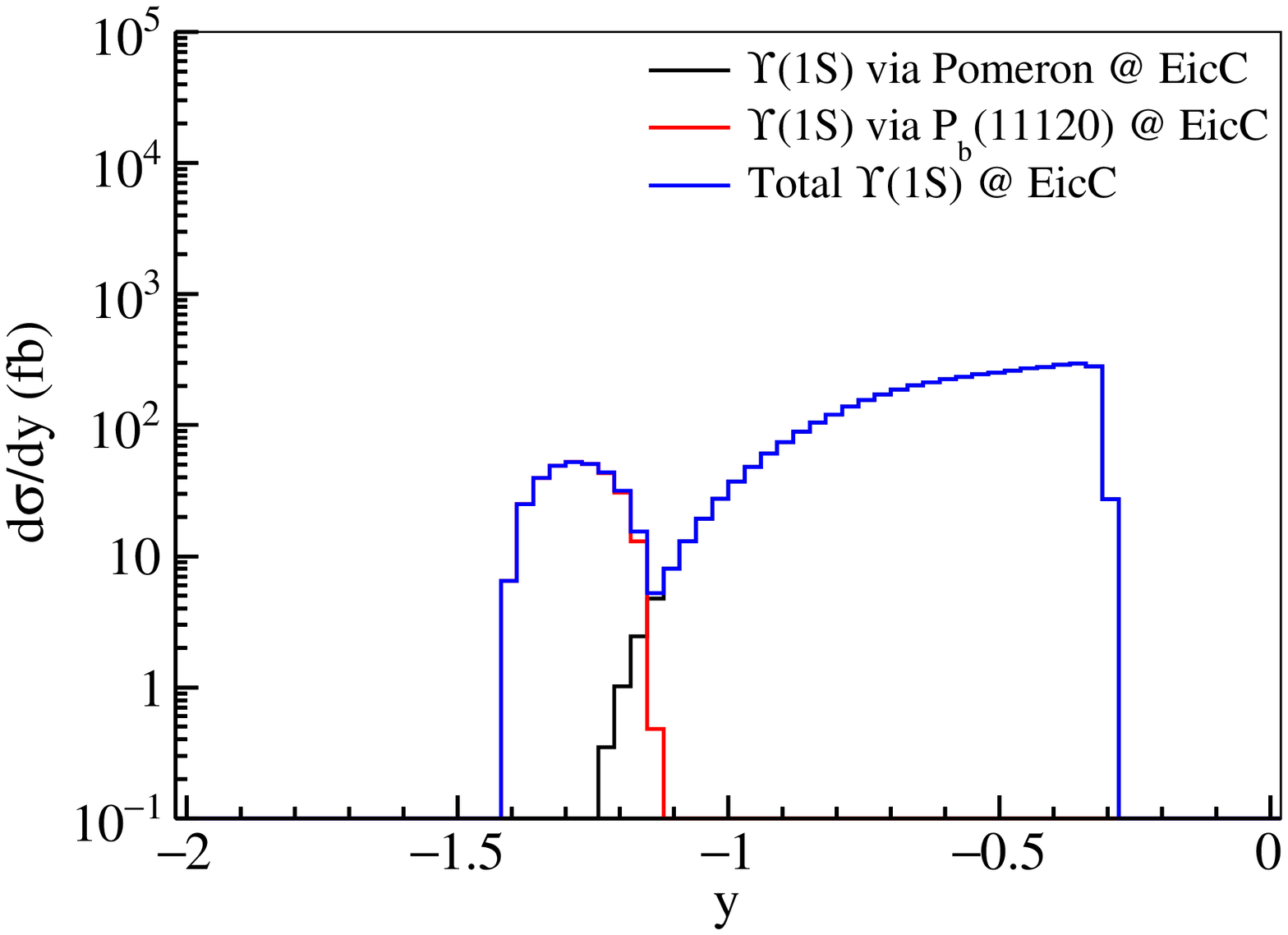}
\includegraphics[width=0.45\textwidth]{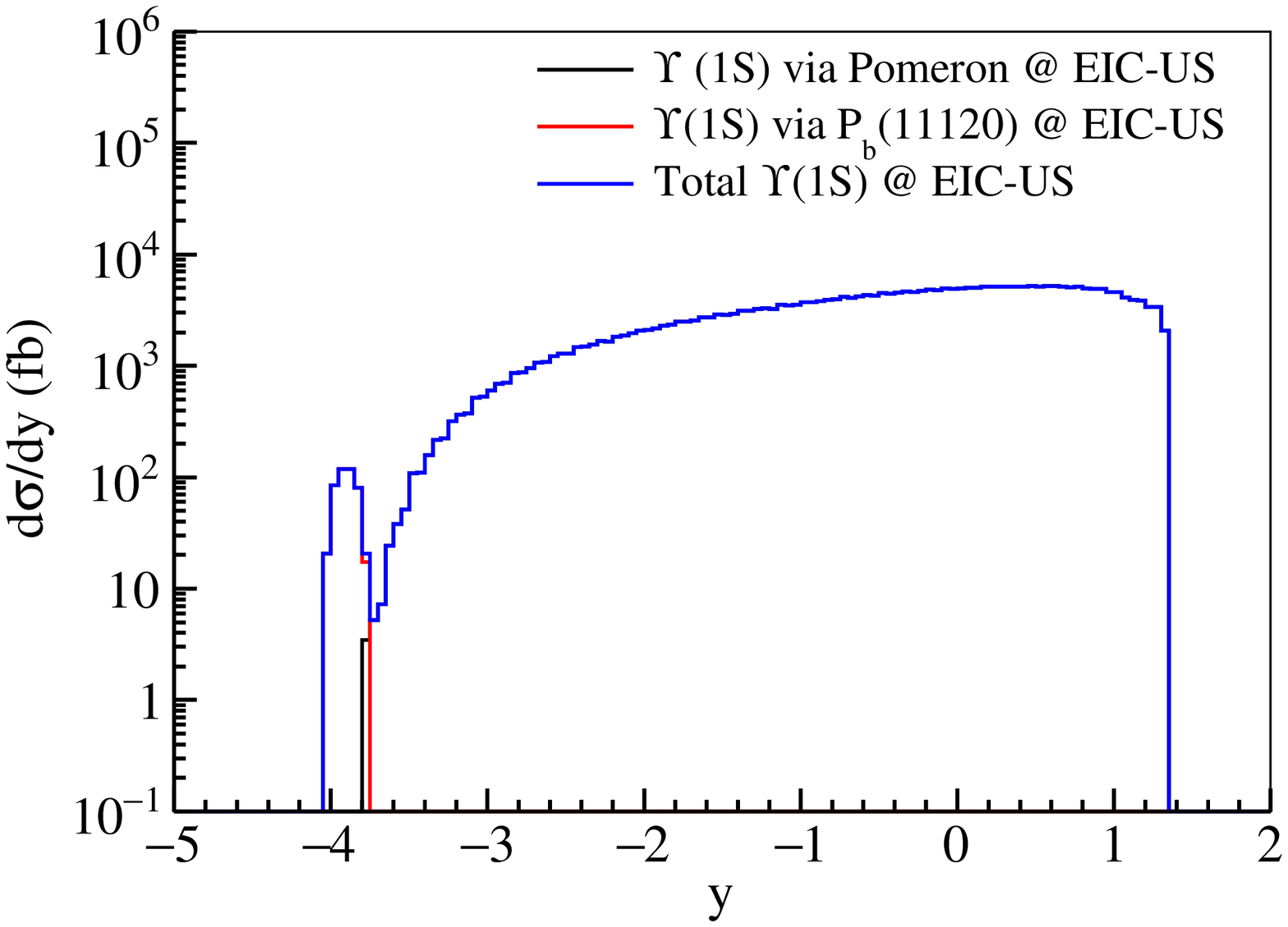}
	\caption{(Color online) Rapidity distributions of $\Upsilon (1S)$ in two channels for proposed EicC (left graph) and EIC-US (right graph). The
		width of $P_b(11120)$ 30 MeV is taken in the calculations.}
	\label{fig05}
\end{figure}

\begin{figure}[h]
	\centering
	\includegraphics[width=0.45\textwidth]{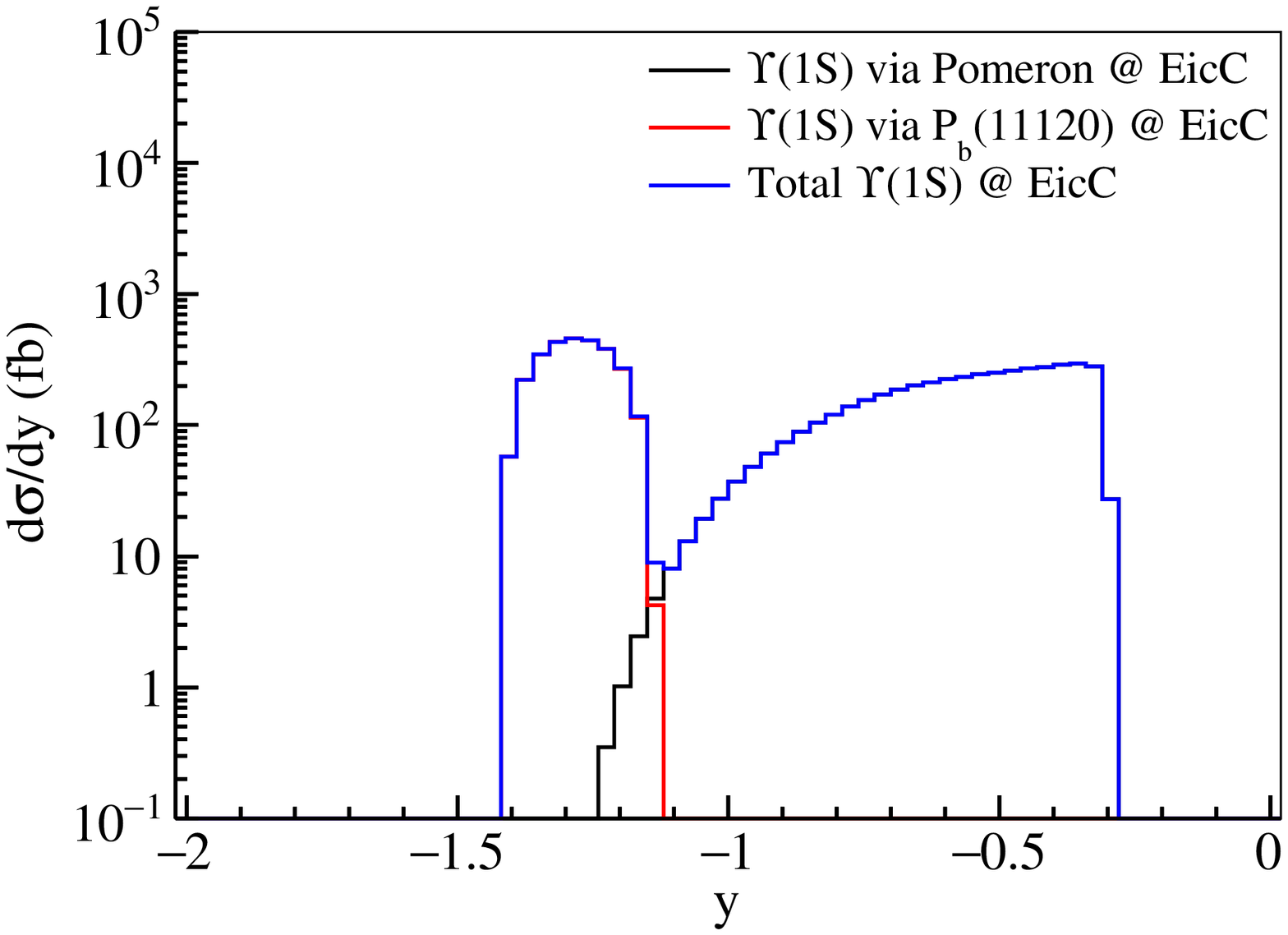}
	\includegraphics[width=0.45\textwidth]{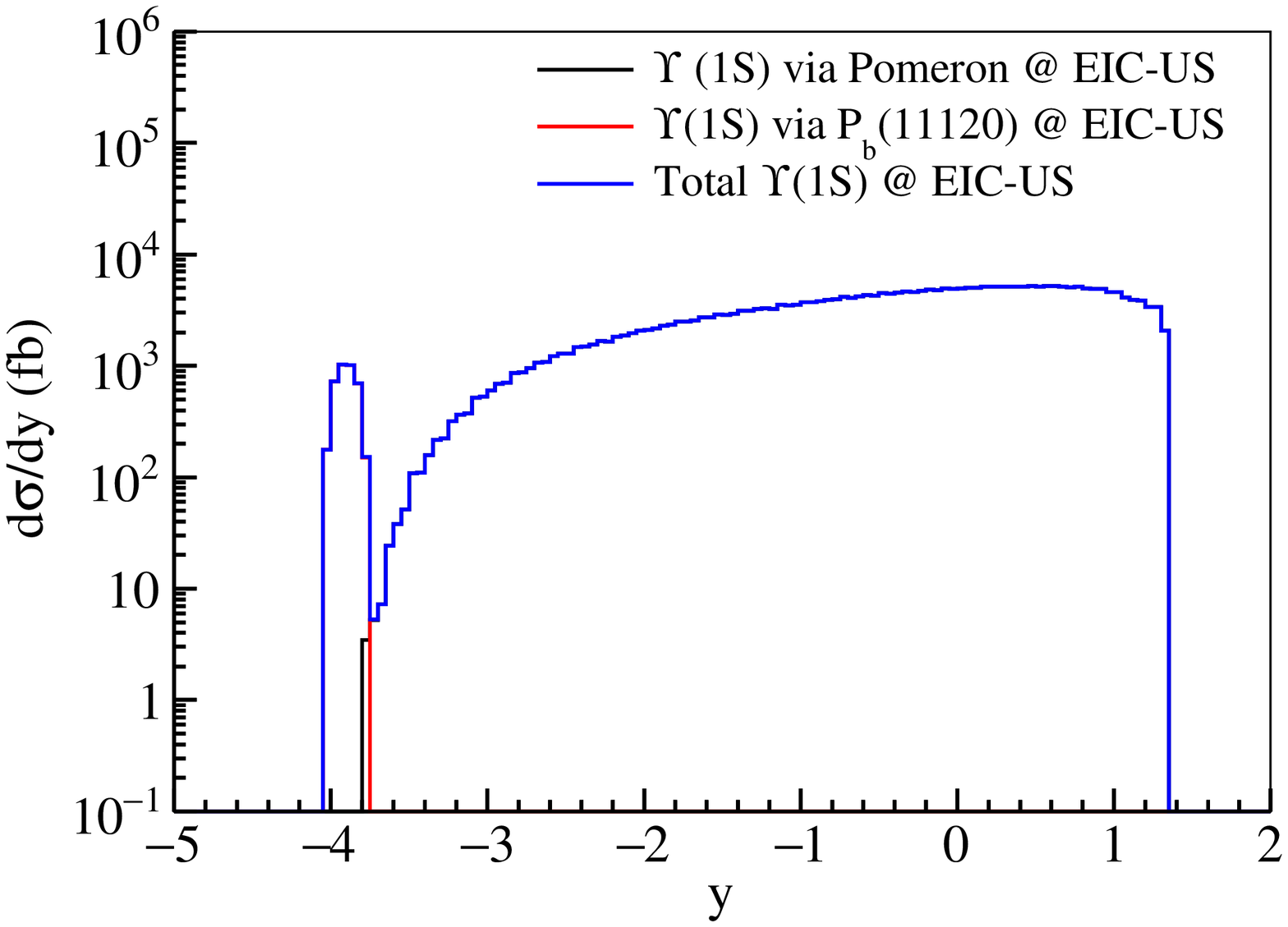}
	\caption{(Color online) Rapidity distributions of $\Upsilon (1S)$ in two channels for proposed EicC (left graph) and EIC-US (right graph). The
		width of $P_b(11120)$ 300 MeV is taken in the calculations.}
	\label{fig06}
\end{figure}

The rapidity distributions of $J/\psi$ in the two channels for proposed EicC and EIC-US are depicted in Fig.~\ref{fig02}. It indicates that $s$-channel is too weak to identify the pentaquark states in rapidity distributions. From Fig.~\ref{fig01} and Fig.~\ref{fig02}, it can be seen that it is difficult to distinguish the contributions from pentaquark resonance
channel as the background. It is difficult to identify the pentaquark states in $J/\psi$ +$p$ production. \\
\indent Moreover, the distributions of $\Upsilon(1S)$ are shown in Fig.~\ref{fig03} to Fig.~\ref{fig06}. Because the width of $P_b(11120)$ are not
determined now, we use 30 -- 300 MeV for the range of width for it \cite{Cao:2019gqo}. In Fig.~\ref{fig03}, the pseudo-rapidity distributions
of $\Upsilon(1S)$ are shown in two channels with lower limit of width. The upper limit of $P_b(11120)$ are applied for the calculations and 
the results are depicted in Fig.\ref{fig04}. From FIg.~\ref{fig03} and Fig.~\ref{fig04}, it can be seen that the peak of $\Upsilon(1S)$ in pentaquark resonance exchange channel is remarkable comparing to the background of the pomeron exchange channel, especially in EicC. The reason is that
the cross section of $\Upsilon(1S)$ in the $t$-channel in EicC is much smaller than the cross section in EIC-US as listed in Table.~\ref{table01}.

Furthermore, the rapidity distributions of $\Upsilon(1S)$ in two channels in lower and upper limit of width of $P_b(11120)$ are presented in
Fig.~\ref{fig05} and Fig.~\ref{fig06}. The same conclusions can be concluded from the rapidity distributions comparing to the pseudo-rapidity
distributions. These results indicate that the $P_b$ pentaquark states of EicC are produced near mid-rapidity region, however, the $P_b$ pentaquark state are produced at large rapidity region at EIC-US because the collider energies of EIC-US is much higher than EicC. Hence, it is easy to identify
$P_b$ states in EicC platform since the detector system can be observe the $P_b$ easily at mid-rapidity region.   \\
\indent Finally, from above discussions, it can be concluded that $P_c(4312)$ is difficult to identify in electron-proton scattering process in proposed EicC and EIC-US because
the strong background of the $t$-channel. On the other hand, the signals of $P_b(11120)$ are remarkable in electron-proton scattering,
especially in proposed EicC. EicC will be a good platform to search $P_b$ pentaquark states in the future according prediction in this paper.
\section{conclusion} \label{sec:conclusion}
In this paper, the hidden-charm and hidden-bottom pentaquark states have been investigated via photoproduction in electron-proton scattering.
The pseudo-rapidity distributions and rapidity distributions of two vector mesons for EicC and EIC-US are compared here under various energy configuration. The $P_c(4312)$ pentaquark resonance state is difficult to identify via pseudo-rapidity distributions in EicC and EIC-US. 
 It can conclude that the $P_b(11120)$ resonance state can be identify via pseudo-rapidity distributions in EicC and EIC-US. EicC is a good platform
 to study the $P_b$ pentaquark resonance states. 
 
Generally speaking, we find that the production cross sections increase slowly with the growing c.m. energies of EIC machine. At high-energy colliders like the proposed EIC-US, the final states are produced at far forward rapidity region. For lower energy colliders like EicC, the systems are produced closer to mid-rapidity region, it is easy to detect the final states by the central detectors.
Our study is a good start point to further detailed simulation of $P_c$ and $P_b$ electroproduction process, which will be helpful for the design of experimental method and detector system for future EICs.

As the EICs are expected to be in operation in near future and unavailable at present, alternative way at hand would be the ultra-peripheral pA collisions at STAR and ALICE~\cite{Goncalves:2019vvo,Xie:2020wfe}. The vector meson production in heavy ions ultra-peripheral collisions can be simulated by STARlight package~\cite{Klein:2016yzr} and the production of pentaquark can be included by a similar extension of kinematic condition in this paper.

\section*{Acknowledgment}
The authors thank gratefully to the discussions with  Dr. J. J. Xie, Dr. Z. Yang, Dr. X.Y.Wang and Dr. J. J. Wu. The work is supported by the National Natural Science Foundation of China (Grant Nos. 11975278, 11405222), and by the Key Research Program of the Chinese Academy of Sciences (Grant NO. XDPB09).

\end{document}